\documentclass{ws-procs975x65}

\def\beq{\begin{equation}}
\def\eeq{\end{equation}}

\begin{document}

\title{Novel stability approach of thin-shell gravastars}

\author{Francisco S.~N.~Lobo} 

\address{Instituto de Astrof\'isica e Ci\^encias do Espa\c{c}o, Universidade de Lisboa, \\
	Faculdade de Ci\^encias, Campo Grande, PT1749-016 Lisboa, Portugal.
}

\author{Prado Mart\'{i}n-Moruno}

\address{Departamento de F\'{i}sica Te\'{o}rica I, Universidad Complutense de Madrid,\\ E-28040 Madrid,
Spain.}

\author{Nadiezhda Montelongo-Garc\'{i}a}

\address{Departamento de F\'{i}sica,
Centro de Investigaci\'{o}n y de Estudios Avanzados del IPN,
A.P. 14-740, 07000 M\'{e}xico D.F., M\'{e}xico.}

\author{Matt Visser}

\address{School of Mathematics and Statistics, Victoria University of Wellington, \\PO Box 600, Wellington 6140, New Zealand.}

\begin{abstract}
We develop an extremely general and robust framework that can be adapted to wide classes of generic spherically symmetric thin-shell gravastars. The thin shell (transition layer) will be permitted to move freely in the bulk spacetimes, permitting a fully dynamic analysis. This will then allow us to perform a general stability analysis, where it is explicitly shown that stability of the gravastar is related to the properties of the matter residing in the thin-shell transition layer. 
\end{abstract}


\bodymatter

\section{Introduction}

{\it Gravastars} (gravitational-vacuum stars) are hypothetical alternatives to standard Schwarzschild black holes,~\cite{gr-qc/0012094, gr-qc/0109035, gr-qc/0405111, gr-qc/0407075, Visser:2003, Cattoen:2005} wherein an interior nonsingular spacetime, such as the de~Sitter spacetime, is matched to some standard exterior geometry resembling the Schwarzschild geometry. The transition layer is located at a radius slightly greater than the Schwarzschild radius. Thus, the gravastar model has no singularity at the origin and no event horizon.

In this work, we describe an extremely general and robust framework that can be adapted to wide classes of generic thin-shell gravastars.~\cite{MartinMoruno:2011rm} Let us consider standard general relativity, with the transition layer confined to a thin shell. The interior and exterior spacetimes on either side of the transition layer are assumed to be spherically symmetric and static, but otherwise arbitrary. We introduce a novel approach, in the context of  linearized stability analysis, wherein the transition layer may move freely in the bulk spacetimes. This allows one to perform a general stability analysis, where stability is related to the properties of the matter residing in the thin-shell transition layer.

\section{General formalism}

Consider two distinct generic spacetime manifolds, namely, an \emph{exterior} ${\cal M_+}$, and an \emph{interior} ${\cal M_-}$, that are eventually to be joined together across a surface layer $\Sigma$. Let the two bulk spacetimes have metrics given by $g_{\mu \nu}^+(x^{\mu}_+)$ and $g_{\mu \nu}^-(x^{\mu}_-)$, in terms of independently defined coordinate systems $x^{\mu}_+$ and $x^{\mu}_-$.
Now specialize this to the case of  two generic static spherically symmetric spacetimes given by the following line elements:
\begin{eqnarray}
\hspace{-1.0cm}ds^2 = - e^{2\Phi_{\pm}(r_{\pm})}\left[1-\frac{b_{\pm}(r_{\pm})}
{r_{\pm}}\right] dt_{\pm}^2 + 
\left[1-\frac{b_{\pm}(r_{\pm})}{r_{\pm}}\right]^{-1}\,dr_{\pm}^2 + r_{\pm}^2 
d\Omega_{\pm}^{2}.
\label{generalmetric}
\end{eqnarray}
Here $\pm$ refers to the exterior and interior geometry, respectively.


The interior and exterior manifolds are bounded by isometric hypersurfaces $\Sigma_+$ and $\Sigma_-$,  with induced metrics $g_{ij}^+$ and $g_{ij}^-$. By assumption $g_{ij}^+(\xi)=g_{ij}^-(\xi)=g_{ij}(\xi)$, with natural hypersurface coordinates $\xi^i=(\tau, \theta, \phi)$. A single manifold ${\cal M}$ is obtained by gluing together ${\cal M_+}$ and ${\cal M_-}$ at their boundaries.~\cite{Israel} This implies that ${\cal M}={\cal M_+}\cup {\cal M_-}$, with the natural identification of the boundaries $\Sigma=\Sigma_+=\Sigma_-$. The intrinsic metric on $\Sigma$ is given by $ds^2_{\Sigma}=-d\tau^2 + a(\tau)^2 \,(d\theta ^2+\sin^2{\theta}\,d\phi^2)$.

Now, to determine the surface stresses that reside on the thin shell, we use the Lanczos equations,~\cite{Israel} which follow from the Einstein equations applied to the hypersurface joining the bulk spacetimes, and are given by
\begin{equation}
S^{i}_{\;j}=-\frac{1}{8\pi}\,\left(\kappa ^{i}_{\;j}-\delta^{i}_{\;j}\;\kappa ^{k}_{\;k}\right)  \,.
\end{equation}
Here $S^{i}_{\;j}$ is the surface stress-energy tensor on $\Sigma$, while the discontinuity of the extrinsic curvature $K_{ij}^{\pm}$ is defined as $\kappa_{ij}=K_{ij}^{+}-K_{ij}^{-}$.
In particular, due to spherical symmetry, considerable simplifications occur, so that $S^{i}_{\;j}$ may be written in terms of the surface energy density, 
$\sigma$, and the surface pressure, ${\cal P}$, which are given by
\begin{eqnarray}
\sigma&=&-\frac{1}{4\pi a}\left[
 \sqrt{1-\frac{b_{+}(a)}{a}+\dot{a}^{2}}
-\sqrt{1-\frac{b_{-}(a)}{a}+\dot{a}^{2}}
\right],
\label{gen-surfenergy2}
\\
{\cal P}&=&\frac{1}{8\pi a}\left[
\frac{1+\dot{a}^2+a\ddot{a}-\frac{b_+(a)+ab'_+(a)}{2a}}{\sqrt{1-\frac{b_{+}(a)}
{a}+\dot{a}^{2}}}     
+
\sqrt{1-\frac{b_{+}(a)}{a}+\dot{a}^{2}} \; a\Phi'_{+}(a)
\right. \nonumber\\
&&
\quad\;\;\;
\left. -
\frac{1+\dot{a}^2+a\ddot{a}-\frac{b_-(a)+ab'_-(a)}{2a}}{\sqrt{1-\frac{b_{-}(a)}
{a}+\dot{a}^{2}}}
-
\sqrt{1-\frac{b_{-}(a)}{a}+\dot{a}^{2}} \; a\Phi'_{-}(a)
\right],
\label{gen-surfpressure2}
\end{eqnarray}
respectively. The surface mass of the thin shell is given by $m_s(a)=4\pi a^2\sigma$.

A fundamental relation is the conservation identity: $S^{i}_{\;j|i}=\left[T_{\mu \nu}\; e^{\mu}_{\;(j)}n^{\nu}\right]^+_-$. The latter follows from the first and second contracted Gauss--Codazzi equations, together with the Lanczos equations,~\cite{Israel} where the convention $\left[X \right]^+_-\equiv X^+|_{\Sigma}-X^-|_{\Sigma}$ is used.

When interpreting this conservation identity, consider first 
the net discontinuity in the momentum flux which impinges on the shell, and which can be written as $\left[T_{\mu\nu}\; e^{\mu}_{\;(\tau)}\,n^{\nu}\right]^+_-
=\dot{a}\,\Xi$. Here we have defined the useful flux term $\Xi$ given by
\begin{equation}
\Xi =\frac{1}{4\pi a}\, \left[
\Phi_+'(a)\sqrt{1-\frac{b_+(a)}{a}+\dot{a}^{2}}
- 
\Phi_-'(a)\sqrt{1-\frac{b_-(a)}{a}+\dot{a}^{2}}
\right]\,.
\end{equation}
Thus, in the general case, taking into account the surface area of the thin shell, i.e., $A=4\pi a^2$, the conservation identity provides:
\begin{equation}
\frac{d(\sigma A)}{d\tau}+{\cal P}\,\frac{dA}{d\tau}=\Xi \,A \, \dot{a}\,.
\label{E:conservation2}
\end{equation}
The first term is the the variation of the internal energy of the shell; the second term is the work done by the shell's internal force; and third term represents the work done by the external forces. Note that the conservation equation can also be written as $\sigma'=-2\,(\sigma +{\cal P})/a+\Xi$, where $ \sigma'=d\sigma /da$. 

In particular, in situations of vanishing flux $\Xi=0$ one obtains the so-called  ``transparency condition'',~\cite{Ishak,Garcia:2011aa} namely $\left[G_{\mu\nu}\; U^{\mu}\,n^{\nu}\right]^+_-=0$. But in general situations this ``transparency condition'' does not hold, and one needs the full version of the conservation equation.

\section{Linearised stability analysis}
\subsection{Equation of motion}

To analyze the stability of the thin shell, it is useful to rearrange the surface energy density $\sigma(a)$ into the form ${1\over2}\dot{a}^2+V(a)=0$, where the potential is given by
\begin{equation}
V(a)= {1\over2}\left\{ 1-{\bar b(a)\over a} -\left[\frac{m_{s}(a)}{2a}\right]^2-\left[\frac{\Delta(a)}{m_{s}
(a)}\right]^2\right\}\,.
   \label{potential}
\end{equation}
The quantities $\bar b(a)$ and $\Delta(a)$ are defined, for simplicity, as
\begin{eqnarray}
\bar b(a)=\frac{b_{+}(a)+b_{-}(a)}{2},\qquad
\Delta(a)=\frac{b_{+}(a)-b_{-}(a)}{2}. \nonumber
\end{eqnarray}
Note that $V(a)$ is a function of the surface mass $m_s(a)$. It is sometimes useful to reverse the logic flow and determine the surface mass as a function of the potential:
	\begin{equation}
	m_s(a) = -a\left[ \sqrt{ 1- {b_+(a)\over a} - 2V(a)} -\sqrt{ 1- {b_-(a)\over a} - 
	2V(a)} \right].
	\end{equation}
Note the logic: By specifying $V(a)$, this tells us how much surface mass is needed on the transition layer, which is implicitly making demands on the equation of state of the matter residing on the transition layer.

Thus, after imposing the equation of motion for the shell, the relevant quantities as functions of $V(a)$, are given by
\begin{eqnarray}
\sigma &=&-\frac{1}{4\pi a}\left[\sqrt{1-\frac{b_{+}}{a} - 2 V}-\sqrt{1-\frac{b_{-}}{a}- 2 V}\right],
\label{gen-surfenergy2-onshell} \\
%
{\cal P}&=&\frac{1}{8\pi a}\left[
\frac{1-2V-aV'-\frac{b_+ +ab'_+}{2a}}{\sqrt{1-\frac{b_{+}}{a}-2V}}     
+
\sqrt{1-\frac{b_{+}}{a}-2V} \; a\Phi'_{+}
\right. \nonumber\\
&&
\qquad
\left. -
\frac{1-2V-aV'-\frac{b_- +ab'_-}{2a}}{\sqrt{1-\frac{b_{-}}{a}-2V}}
-
\sqrt{1-\frac{b_{-}}{a}-2V} \; a\Phi'_{-}
\right],
\label{gen-surfpressure2-onshell} \\
%
\hspace{-0.75cm}
\Xi &=&\frac{1}{4\pi a}\, \left[\Phi'_+ \sqrt{1-\frac{b_+}{a}-2V} - \Phi'_- \sqrt{1-\frac{b_-}{a}-2V}\right]\,,
\end{eqnarray}
respectively.
These three quantities $\{\sigma(a),{\cal P}(a),\Xi(a)\}$ are inter-related by the conservation law, so at most two of them are independent.

Consider a linearization around an assumed static solution, $a_0$, and a Taylor expansion of $V(a)$ around $a_0$ to second order. Expanding around a static solution $\dot a_0=\ddot a_0 = 0$, we have 
$V(a_0)=V'(a_0)=0$, so it is sufficient to consider
\begin{equation}
V(a)= \frac{1}{2}V''(a_0)(a-a_0)^2+O[(a-a_0)^3]
\,.   \label{linear-potential}
\end{equation}
The assumed static solution at $a_0$ is stable if and only if $V(a)$ has a local minimum at $a_0$, which requires $V''(a_{0})>0$.

The primary criterion for the thin shell stability is the condition $V''(a_{0})>0$, though it will be useful to rephrase it in terms of more basic quantities. For instance, it is useful to express $m_s'(a)$ and $m_s''(a)$, which allows us to easily study linearized stability, and to develop a simple inequality on $m_s''(a_0)$ using the constraint $V''(a_0)>0$.

In view of the redundancies coming from the relations $m_s(a) = 4\pi\sigma(a) a^2$ and the differential conservation law, the only interesting quantities are  $\Xi'(a)$, $\Xi''(a)$. The relevant quantities to evaluate, at the assumed stable solution $a_0$, are given by $m_s''(a)$ and $\Xi''(a)$.

\subsection{Master equations}

The stability condition $V''(a_0)\geq0$ can be translated into an explicit inequality on $m_s''(a_0)$, given by:
\begin{eqnarray}
 m_s''(a_0) &\geq&
+{1\over4 a_0^3} 
\left\{ 
{ [b_+(a_0)- a_0 b_+'(a_0)]^2\over[1-b_+(a_0)/a_0]^{3/2}} 
- 
{ [b_-(a_0)- a_0 b_-'(a_0)]^2\over[1-b_-(a_0)/a_0]^{3/2}}
\right\}
\nonumber\\
&& 
+{1\over2} 
\left\{ 
{b_+''(a_0)\over\sqrt{1-b_+(a_0)/a_0}} 
-
{b_-''(a_0)\over\sqrt{1-b_-(a_0)/a_0}} \right\},
  \label{stable_ddms1}
\end{eqnarray}
provided $b_+(a_0)\geq b_-(a_0)$, which is equivalent to 
$\sigma(a_0)\geq 0$.
If $b_+(a_0)\leq b_-(a_0)$ the direction of the inequality is reversed. In the absence of external forces this inequality is the only stability constraint one requires.

However, once one has external forces ($\Xi\neq 0$), a second stability condition is imposed:
\begin{eqnarray}
\left.[4\pi\,\Xi(a)\,a]''\right|_{a_0} &\leq& \left.\left\{ 
\Phi_+'''(a) \sqrt{1-b_+(a)/a} - 
\Phi_-'''(a) \sqrt{1-b_-(a)/a} \right\}\right|_{a_0}
\nonumber\\
&& 
- \left.\left\{ 
\Phi_+''(a) { (b_+(a)/a)'\over\sqrt{1-b_+(a)/a}} - 
\Phi_-''(a){(b_-(a)/a)'\over\sqrt{1-b_-(a)/a}} \right\}\right|_{a_0}
\nonumber\\
&&
-{1\over4} 
\left.\left\{ 
\Phi_+'(a) { [(b_+(a)/a)']^2\over[1-b_+(a)/a]^{3/2}} -
\Phi_-'(a) {[(b_-(a)/a)']^2\over[1-b_-(a)/a]^{3/2}} \right\}\right|_{a_0}
\nonumber\\
&&
-{1\over2} 
\left.\left\{ 
\Phi_+'(a) { (b_+(a)/a)''\over\sqrt{1-b_+(a)/a}} -
\Phi_-'(a) {(b_-(a)/a)''\over\sqrt{1-b_-(a)/a}} \right\}\right|_{a_0},
   \label{stability_Xi}
\end{eqnarray}
provided $\Phi'_+(a_0)/\sqrt{1-b_+(a_0)/a_0} \geq \Phi'_-(a_0)/\sqrt{1-b_-(a_0)/a_0} $. Note that this last equation is entirely vacuous in the absence of external forces, which is why it has not appeared in the literature until now.~\cite{MartinMoruno:2011rm,Garcia:2011aa}

\subsection{Specific gravastar model}

In discussing specific gravastar models one now ``merely'' needs to apply the general  formalism described above. Up to this stage we have kept the formalism as general as possible with a view to future applications. However, for pedagogical purposes, we now analyse the traditional gravastar picture.  Namely, we consider a  Schwarzschild exterior (with $b_+(r)=2M$ and 	$\Phi_+(r)=0$) and de~Sitter interior (with $b_-(r) = r^3/R^2$ and $\Phi_-(r)=0$). 

The transition layer is located at $2M < a < R$. (i.e., outside the Schwarzschild event horizon, and inside de Sitter cosmological horizon). The external forces vanish ($\Xi=0$), as $\Phi_\pm =0$.
The inequality one derives for $m_s''(a_0)$ is given by
\begin{eqnarray}
\hspace{-1cm}
a_0\, m_s''(a_0) &\geq&
{ \left(M/a_0\right)^2\over[1-2M/a_0]^{3/2}} 
- 
{ \left(a_0/R\right)^2 \; \left[3-2\left(a_0/R\right)^2\right] \over[1-\left(a_0/R\right)^2]^{3/2}}. 
\end{eqnarray}

\section{Conclusions}

We have developed an extremely general and robust framework leading to the linearized stability analysis of dynamical spherically symmetric thin-shells, and in this report have applied it to the gravastar model. Due to the definition of the normals on the junction interface, a few strategic sign flips arise when comparing thin-shell gravastars~\cite{MartinMoruno:2011rm} and traversable wormhole~\cite{Garcia:2011aa} configurations. It is interesting to note that the surface energy density is always negative for the thin-shell traversable wormholes~\cite{Garcia:2011aa}. We have also shown that an analysis of the conservation law of the surface stresses implies the presence of a flux term, corresponding to the net discontinuity in the (bulk) momentum flux which impinges on the shell.

In the context of the linearized stability analysis we introduced a novel approach, where we have considered the surface mass as a function of the potential, so that specifying the latter tells us how much surface mass we need to put on the transition layer. This procedure demonstrates in full generality that the stability of the thin shell is equivalent to choosing suitable properties for the material residing on the thin shell. We have also considered an applications to the traditional gravastar. Further afield, we expect that the mathematical formalism developed herein will also prove useful when considering spacetime ``voids'' (manifolds with boundary).

\section*{Acknowledgments}
This work was partially supported by the Funda\c{c}\~{a}o para a Ci\^{e}ncia e Tecnologia (FCT) through the grants EXPL/FIS-AST/1608/2013 and UID/FIS/04434/2013. \\
PMM acknowledges financial support from the Spanish Ministry of Economy and Competitiveness (MINECO) through the postdoctoral training contract FPDI-2013-16161, and the project FIS2014-52837-P. \\
FSNL was supported by a FCT Research contract, with reference IF/00859/2012.\\
MV was supported by a James Cook fellowship, and by the Marsden fund, both administered by the Royal Society of New Zealand.

\end{document}